\documentclass[twocolumn]{aastex631}

\usepackage{natbib}
\usepackage{comment}
\usepackage{float}
\usepackage{wasysym}
\usepackage{placeins}
\usepackage{soul}
\usepackage{xcolor}

\newcommand{\hii}{$\mathrm{H}\,${\small II}}

\begin{document}

\title{Probing Shocked Ejecta in SN 1987A: A novel diagnostic approach using XRISM$-$Resolve}

\author[0000-0002-6045-136X]{Vincenzo Sapienza}
\affiliation{Dipartimento di Fisica e Chimica E. Segr\`e, Universit\`a degli Studi di Palermo, Piazza del Parlamento 1, 90134, Palermo, Italy}
\affiliation{INAF-Osservatorio Astronomico di Palermo, Piazza del Parlamento 1, 90134, Palermo, Italy}
\affiliation{Department of Physics, Graduate School of Science, The University of Tokyo, 7-3-1 Hongo, Bunkyo-ku, Tokyo 113-0033, Japan}

\author[0000-0003-0876-8391]{Marco Miceli}
\affiliation{Dipartimento di Fisica e Chimica E. Segr\`e, Universit\`a degli Studi di Palermo, Piazza del Parlamento 1, 90134, Palermo, Italy}
\affiliation{INAF-Osservatorio Astronomico di Palermo, Piazza del Parlamento 1, 90134, Palermo, Italy}

\author[0000-0003-0890-4920]{Aya Bamba}
\affiliation{Department of Physics, Graduate School of Science, The University of Tokyo, 7-3-1 Hongo, Bunkyo-ku, Tokyo 113-0033, Japan}
\affiliation{Research Center for the Early Universe, School of Science, The University of Tokyo, 7-3-1 Hongo, Bunkyo-ku, Tokyo 113-0033, Japan}
\affiliation{Trans-Scale Quantum Science Institute, The University of Tokyo\\ 7-3-1 Hongo, Bunkyo-ku, Tokyo 113-0033, Japan}

\author[0000-0003-2836-540X]{Salvatore Orlando}
\affiliation{INAF-Osservatorio Astronomico di Palermo, Piazza del Parlamento 1, 90134, Palermo, Italy}

\author[0000-0002-2899-4241]{Shiu-Hang Lee}
\affiliation{Department of Astronomy, Kyoto University Oiwake-cho, Kitashirakawa, Sakyo-ku, Kyoto 606-8502, Japan}
\affiliation{Kavli Institute for the Physics and Mathematics of the Universe (WPI), The University of Tokyo, Kashiwa 277-8583, Japan}

\author[0000-0002-7025-284X]{Shigehiro Nagataki}
\affiliation{RIKEN Interdisciplinary Theoretical and Mathematical Sciences Program (iTHEMS), 2-1 Hirosawa, Wako, Saitama 351-0198, Japan}
\affiliation{ Astrophysical Big Bang Laboratory (ABBL), RIKEN Cluster for Pioneering Research, 2-1 Hirosawa, Wako, Saitama 351-0198, Japan}
\affiliation{Astrophysical Big Bang Group (ABBG), Okinawa Institute of Science and Technology Graduate University, 1919-1 Tancha, Onna-son, Kunigami-gun, Okinawa 904-0495, Japan}

\author[0000-0002-0603-918X]{Masaomi Ono}
\affiliation{RIKEN Interdisciplinary Theoretical and Mathematical Sciences Program (iTHEMS), 2-1 Hirosawa, Wako, Saitama 351-0198, Japan}
\affiliation{ Astrophysical Big Bang Laboratory (ABBL), RIKEN Cluster for Pioneering Research, 2-1 Hirosawa, Wako, Saitama 351-0198, Japan}
\affiliation{Institute of Astronomy and Astrophysics, Academia Sinica, Taipei 10617, Taiwan}

\author[0000-0002-1104-7205]{Satoru Katsuda}
\affiliation{Graduate School of Science and Engineering, Saitama University, 255 Simo-Ohkubo, Sakura-ku, Saitama city, Saitama, 338-8570, Japan}

\author{Koji Mori}
\affiliation{Faculty of Engineering, University of Miyazaki, 1-1 Gakuen Kibanadai Nishi, Miyazaki, Miyazaki 889-2192, Japan}
\affiliation{Japan Aerospace Exploration Agency, Institute of Space and Astronautical Science, 3-1-1 Yoshino-dai, Chuo-ku, Sagamihara, Kanagawa 252-5210, Japan}

\author[0000-0003-2008-6887]{Makoto Sawada}
\affiliation{Department of Physics, Rikkyo University, 3-34-1 Nishi Ikebukuro, Toshima-ku, Tokyo 171-8501, Japan}

\author[0000-0002-2359-1857]{Yukikatsu Terada}
\affiliation{Graduate School of Science and Engineering, Saitama University, 255 Simo-Ohkubo, Sakura-ku, Saitama city, Saitama, 338-8570, Japan}

\author[0000-0002-2774-3491]{Roberta Giuffrida}
\affiliation{Dipartimento di Fisica e Chimica E. Segr\`e, Universit\`a degli Studi di Palermo, Piazza del Parlamento 1, 90134, Palermo, Italy}
\affiliation{INAF-Osservatorio Astronomico di Palermo, Piazza del Parlamento 1, 90134, Palermo, Italy}

\author[0000-0002-2321-5616]{Fabrizio Bocchino}
\affiliation{INAF-Osservatorio Astronomico di Palermo, Piazza del Parlamento 1, 90134, Palermo, Italy}



\begin{abstract}
%Context 
Supernova (SN) 1987A is one of the best candidates to exploit the capabilities of the freshly launched XRISM satellite.
This celestial object offers the unique opportunity to study the evolution of a SN into a young supernova remnant.
To date, the X-ray emission has been dominated by the shocked circumstellar medium (CSM), with no shocked ejecta firmly detected. However, 
recent studies provide compelling evidence that in the forthcoming years the X-ray emission from SN 1987A will increasingly stem from the ejecta.
Our aim is to assess the proficiency of XRISM-Resolve high resolution spectrometer in pinpointing signatures of the shocked ejecta in SN 1987A.
Taking advantage of a self consistent state-of-art magneto-hydrodynamic simulation that describes the evolution from SN 1987A to its remnant, we synthesized the XRISM-Resolve spectrum of SN 1987A, as it would be collected in the allocated observation during the performance verification phase, which is foreseen for 2024. 
Our predictions clearly show the leading role of shocked ejecta in shaping the profile of the emission lines. The Doppler broadening associated with the bulk motion along the line of sight of the rapidly expanding ejecta is shown to increase the line widths well above the values observed so far.  
The quantitative comparison between our synthetic spectra and the XRISM spectra will enable us to establish a strong connection between the broadened line emission and the freshly shocked ejecta. This, in turn, will allow us to retrieve the ejecta dynamics and chemical composition from the X-ray emission.
\end{abstract}

%%\keywords{Classical Novae (251) --- Ultraviolet astronomy(1736) --- History of astronomy(1868) --- Interdisciplinary astronomy(804)}


\section{Introduction} \label{sec:intro}
Situated at 51.4 kpc \citep{1999IAUS..190..549P} in the Large Magellanic Cloud (LMC), Supernova (SN) 1987A is the relic of a core collapse SN (CCSN) discovered on February 23rd, 1987  \citep{1987A&A...177L...1W}.
The supernova remnant (SNR) is evolving in a highly inhomogeneous circumstellar medium (CSM).
The CSM is characterized by a dense and clumpy equatorial ring (ER) structure embedded within a diffuse \hii\ region \citep{2005ApJS..159...60S}.
SN 1987A offers, for the first time in modern astronomy, the opportunity of resolving the evolution of a SN into its SNR through all the wavelengths of the electromagnetic spectrum \citep{2016ARA&A..54...19M}, being an \textit{unicum} in this field.

Observations in the X-ray band play a pivotal role in probing the interaction between the shock front and both the CSM and the external envelope of ejecta, as well as in studying shock physics.
Hard ($10-30$ keV) X-ray emission in the region of SN 1987A was first detected six months after the explosion \citep{1987Natur.330..230D}.
Subsequently, ROSAT observed a soft ($0.1-2.4$ keV) X-ray thermal emission $\sim1500$ days after the explosion \citep{1994A&A...281L..45B}.
Since the initial detection, the X-ray emission has continuously increased for about 25 yrs, and its nature has been investigated by several studies, confirming the encounter of the shock with the equatorial ring (\citealt{1997ApJ...477..281B},\citealt{2005ApJ...634L..73P}, \citealt{2006A&A...460..811H}, \citealt{2009ApJ...692.1190Z}, \citealt{2012A&A...548L...3M}).
Recent investigations in the past few years, showed that the soft X-ray emission ($0.5-2$ keV) is decreasing, while the harder component ($3-8$ keV) continues to increase (\citealt{2016ApJ...829...40F}, \citealt{2021ApJ...916...41S}, \citealt{2021ApJ...922..140R}, \citealt{2022A&A...661A..30M}).
Additionally, indications of a Pulsar Wind Nebula (PWN), embedded in the dense and cold innermost ejecta, were found by \citet{2021ApJ...908L..45G,2022ApJ...931..132G} in the $10-20$ keV band.

The overall X-ray light-curve has been impressively replicated by 3D Hydrodynamic (HD) and magneto-hydrodynamic (MHD) simulations by \citet{2015ApJ...810..168O,2019A&A...622A..73O,2020A&A...636A..22O}.
These simulations offer valuable insights into the temporal evolution of the X-ray emission.
They indicate that the X-ray emission was initially dominated by the shocked \hii\ region.
However, as time progressed, a notable transition occurred, with the emission becoming increasingly influenced by the shocked dense ER.
According to the models, the remnant is currently entering a third phase, where the X-ray emission is predominantly driven by the SN outermost ejecta, heated by the reverse shock.
Therefore, X-ray features from the shocked ejecta are expected to be observed in the near future.

In a pioneering study, \citet{2019NatAs...3..236M} introduced a novel data analysis approach for examining the case of SN 1987A.
They analyzed the \textit{Chandra} High Energy Transmission Grating data from 2007 and 2011 comparing them with the 3D HD simulation presented by \citet{2015ApJ...810..168O}.
Since the simulation is able to reproduce self-consistently the whole broadening (due to the thermal motion of ions and to bulk motion along the line of sight) of the spectral lines, they were able to deduce the post shock temperature of protons and ions through the comparison of the model with observation (see also \citealt{2023PPCF...65c4003M}, \citealt{2021ApJ...922..140R}).

The JAXA-NASA collaborative mission X-ray Imaging and Spectroscopy Mission (XRISM, \citealt{2020SPIE11444E..22T}), launched on September 7th 2023, with the Resolve spectrometer \citep{2022SPIE12181E..1SI} will offer non-dispersive, high-resolution X-ray spectroscopy at $<$7 eV energy resolution with a large effective area around 6 keV.
These capabilities open the door for in-depth investigations of young remnants like SN 1987A.
Moreover, the XRISM mission will observe it in the Performance Verification (PV) phase.

Leveraging an in-house developed tool, we synthesize the XRISM-Resolve X-ray spectrum of SN 1987A from the MHD simulations by \citet{2020A&A...636A..22O} and \cite{2020ApJ...888..111O}, which have been proven to accurately reproduce the observables obtained from current telescopes.
We adopt an approach similar to \cite{2019NatAs...3..236M}, and aim at finding new diagnostics to trace ejecta signatures for the forthcoming observations of SN 1987A.

The structure of this letter is as follows:
Section \ref{sect:mod} presents the model and details the spectral synthesis process, while Section \ref{sect:res} showcases the results obtained from the synthesized models. 
Discussions and conclusions are drawn in Section \ref{sect:con}.

\section{Model and synthesis procedure}\label{sect:mod}
For our spectral synthesis, we harness the 3D MHD model by \citet{2020A&A...636A..22O}.
This state-of-the-art simulation models the evolution of SN 1987A (both ejecta and CSM) from the immediate aftermath ($\sim20$ hours) of the 3D core-collapse model provided by \cite{2020ApJ...888..111O}.
We adopted the model configuration that best matches the observations for the progenitor star, the SN, and the SNR (B18.3 in \citealt{2020A&A...636A..22O} and hereafter).
The progenitor star model for this configuration was proposed by \cite{2018MNRAS.473L.101U}.
It implies a 18.3 $M_{\astrosun}$ blue supergiant progenitor, resulting from the slow-merging \citep{2002MNRAS.334..819I} of two massive stars with 14 $M_{\astrosun}$ and 9 $M_{\astrosun}$.
This model can reproduce the luminosity and surface temperature of the progenitor star of SN 1987A.
The evolution from the progenitor to the CCSN was then followed by \cite{2020ApJ...888..111O}, and then evolved into SNR by \cite{2020A&A...636A..22O}.
The B18.3 model reproduces the X-ray light-curves and the evolution of the size and morphology of the X-ray emitting structures of remnant of SN 1987A.
Moreover, it is able to reproduce the evolution of the X-ray spectra \citep{2019NatAs...3..236M}, as well as the observed velocity distributions of $^{44}$Ti \citep{2015Sci...348..670B} and $^{56}$Fe \citep{1990ApJ...360..257H} and the spatial distribution of the SiO and CO molecules \citep{2017ApJ...842L..24A}.
\begin{figure*}
    \centering
    \includegraphics[width=\textwidth]{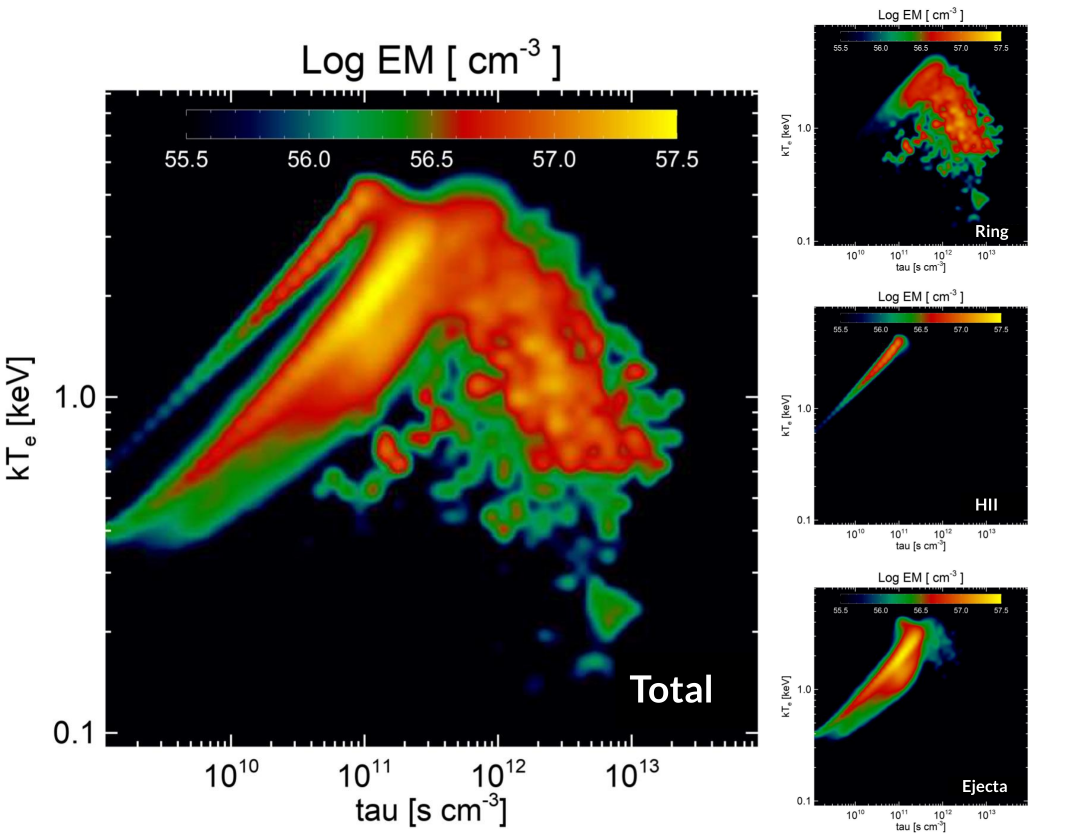}
    \includegraphics[width=\columnwidth]{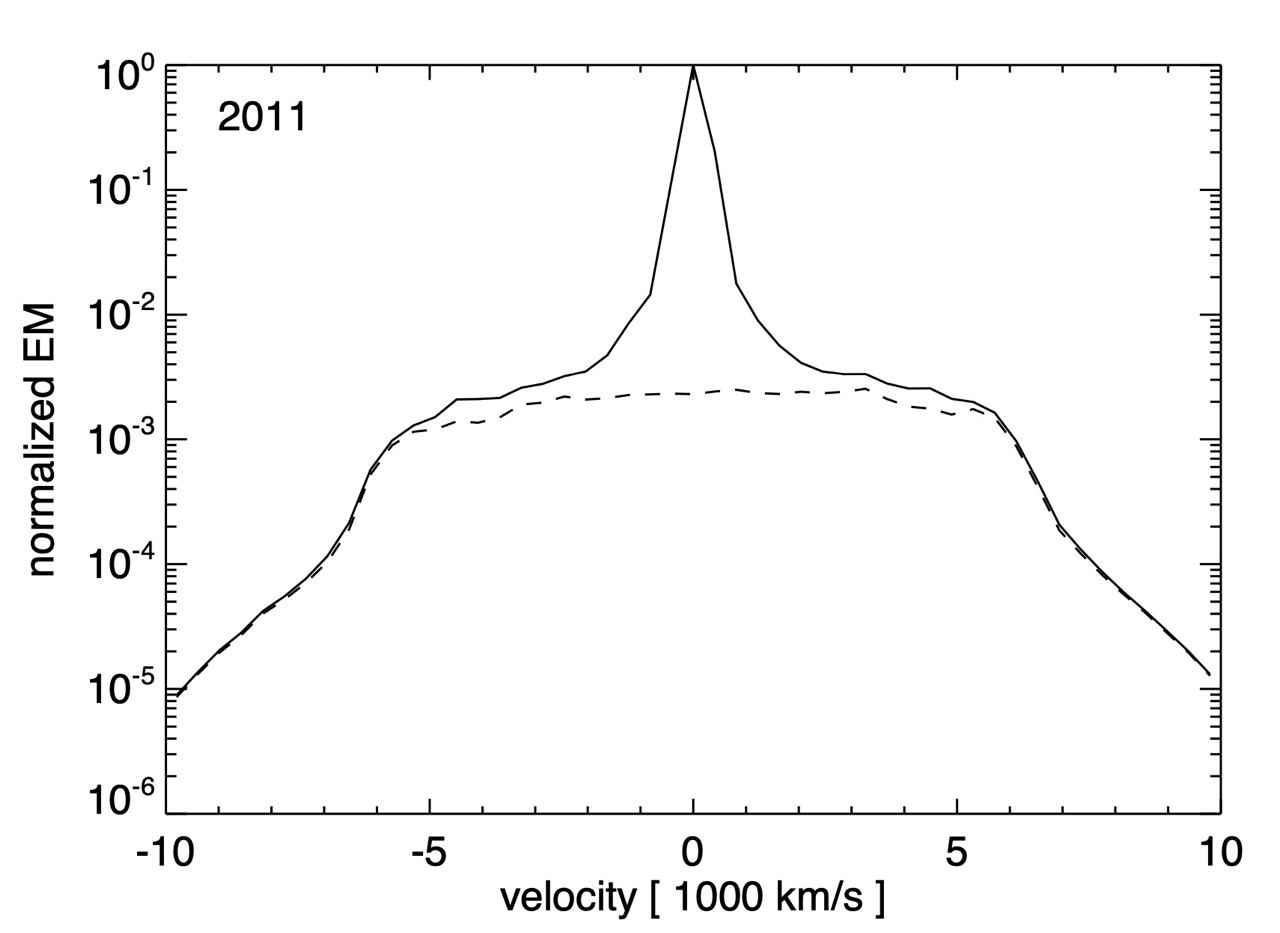}
    \includegraphics[width=\columnwidth]{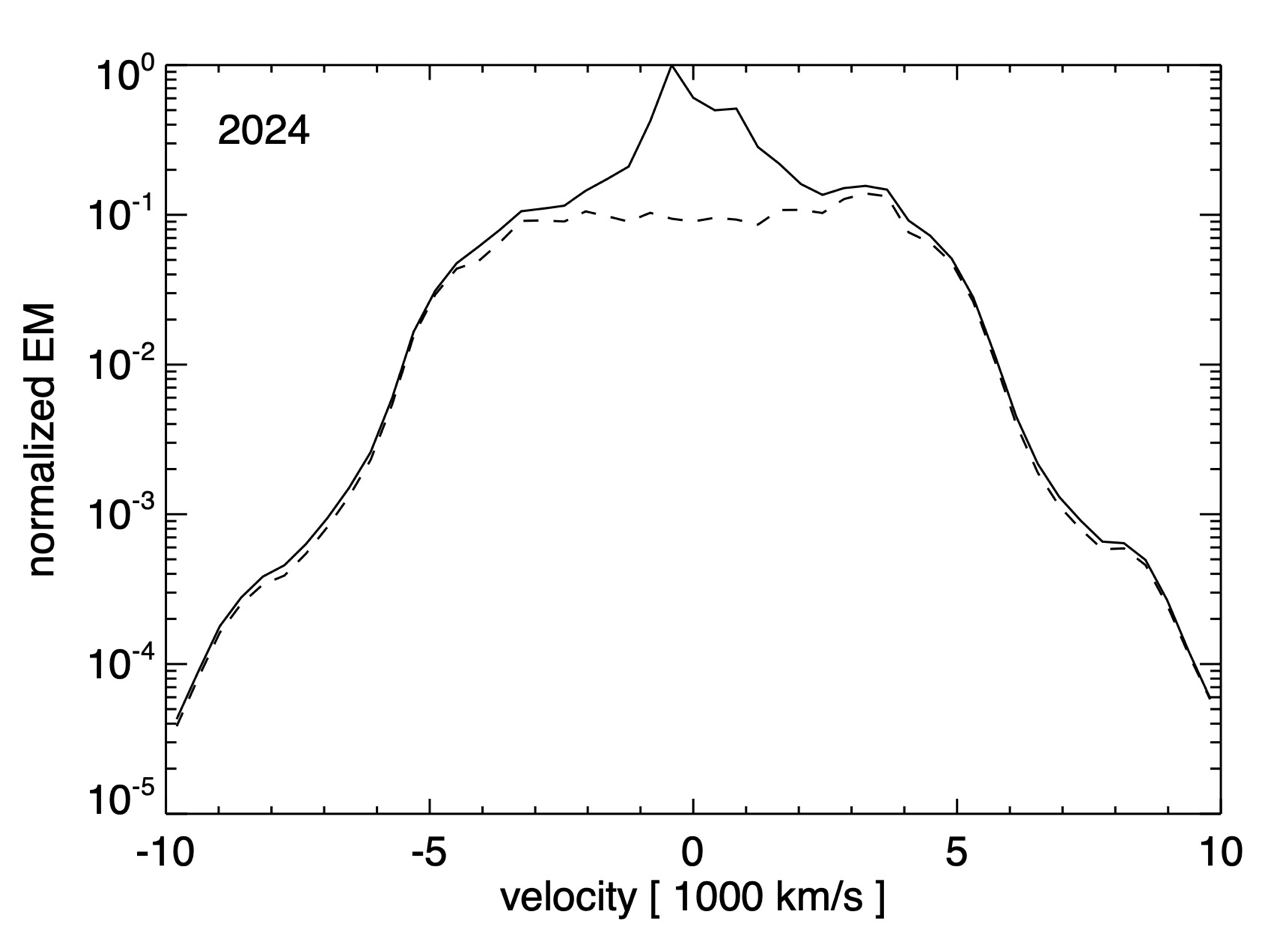}
    \caption{\textit{Upper Left panel:} Distribution of emission measure (EM) for the total X-ray emitting plasma in SN 1987A as a function of the plasma temperature and ionization parameter, as predicted by our MHD model for year 2024.
    \textit{Upper Right panels:} Same as left panel for the ER, \hii\ region and ejecta from top to bottom, respectively.
    \textit{Lower left panel:} Normalized EM distribution as a function of the velocity along the line of sight for year 2011.
    Black solid line marks the total EM, while the black dashed line marks the contribution of the ejecta.
    Negative velocity are towards the observer.
    \textit{Lower right panel}: Same as \textit{lower left panel} for the year 2024}
    \label{fig:ktvtau}
\end{figure*}

Fig. \ref{fig:ktvtau} (upper panels) shows the distribution of the emission measure (EM) of the X-ray emitting plasma as a function of the electron temperature $kT$ and of the ionization parameter $\tau$ (i.e., the time integral of the electron density computed from the impact with the shock front), as predicted by our model of SN 1987A for year 2024.
The panels on the upper-right show the contribution to the EM from the ER, the \hii\ region and the ejecta, respectively from top to bottom.
The contribution to the EM from the ejecta (which are moving faster than the CSM, see lower panels of Fig. \ref{fig:ktvtau}) shows the highest increase (compared to the ER and \hii\ region) with respect to previous epochs (see \citealt{2021ApJ...922..140R,2022ApJ...931..132G}). 
It is also worth noting that the ER shows values of $\tau$ that are, on average, higher than that of the ejecta.

To self-consistently synthesize the thermal X-ray emission from the 3D MHD simulation, we employ an in-house developed tool in a similar fashion as  \cite{2019NatAs...3..236M}, \cite{2020A&A...638A.101G}, and \cite{2023PPCF...65c4003M}. 
In each cell of the computational domain we derive the value of electron and proton temperature, electron density and $\tau$.
We use the aforementioned quantities as input parameters for the non-equilibrium of ionization optically thin plasma model \texttt{vnei} from XSPEC \citep{1996ASPC..101...17A} (version 12.12.0, ATOMDB version 3.0.9).
We include the effects of the interstellar and intergalactic absorption by using the \texttt{tbabs} model in XSPEC, with an equivalent column density $n_H=0.235\times10^{20}$ $\mathrm{cm}^{-2}$ as in \citet{2006ApJ...646.1001P}.
The X-ray emission models, folded through the response matrix with an energy resolution of 7 eV\footnote{We adopted the response matrix available at the XRISM-JAXA Researchers website (\url{https://xrism.isas.jaxa.jp/research/proposer/obsplan/response/index.html})} (High resolution grade only) of XRISM-Resolve, are calculated using XSPEC for different values of electron temperature and ionization time, and stored in a look-up table.

The MHD simulation shows that in 2024 the bulk of shocked ejecta belongs to the outer mantle (which is mainly composed of H and He, \citealt{2020ApJ...888..111O}). 
For the spectral synthesis, we assume that this external envelope of ejecta has the same chemical composition as the CSM. 
In particular, we adopt the abundances of \citet{2009ApJ...692.1190Z}.
Though also a fraction of metal-rich ejecta is expected to be shocked in 2024, we estimate from the simulation that their contribution to the ejecta X-ray emission is less than $0.0001\%$ of the total, so we will not consider it in this paper.

Our tool incorporates the ability to add in the spectral synthesis the line broadening arising from the bulk velocities along the line of sight of the different X-ray emitting parts of the remnant, namely: ejecta and CSM (ER and \hii\ region).
In addition to this effect, we also include the thermal broadening arising from the thermal motion of the ions in the plasma.
To this end, we assume the ion temperature to be mass-proportional with respect to the proton temperature (\citealt{2019NatAs...3..236M,2023PPCF...65c4003M}).
Once obtained the complete spectral model by summing cell by cell the computed spectra, we applied a Poissonian randomization to synthesize the spectra.

\section{Results} \label{sect:res}
\begin{figure*}[ht!b]
    \centering
    \includegraphics[width=\textwidth]{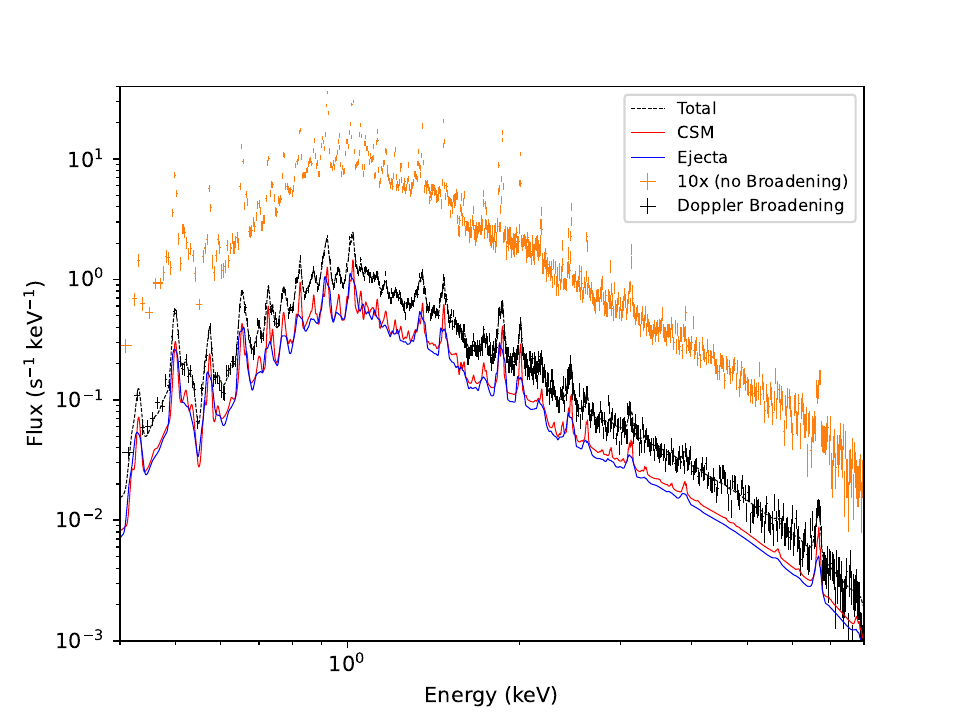}
    \includegraphics[width=\columnwidth]{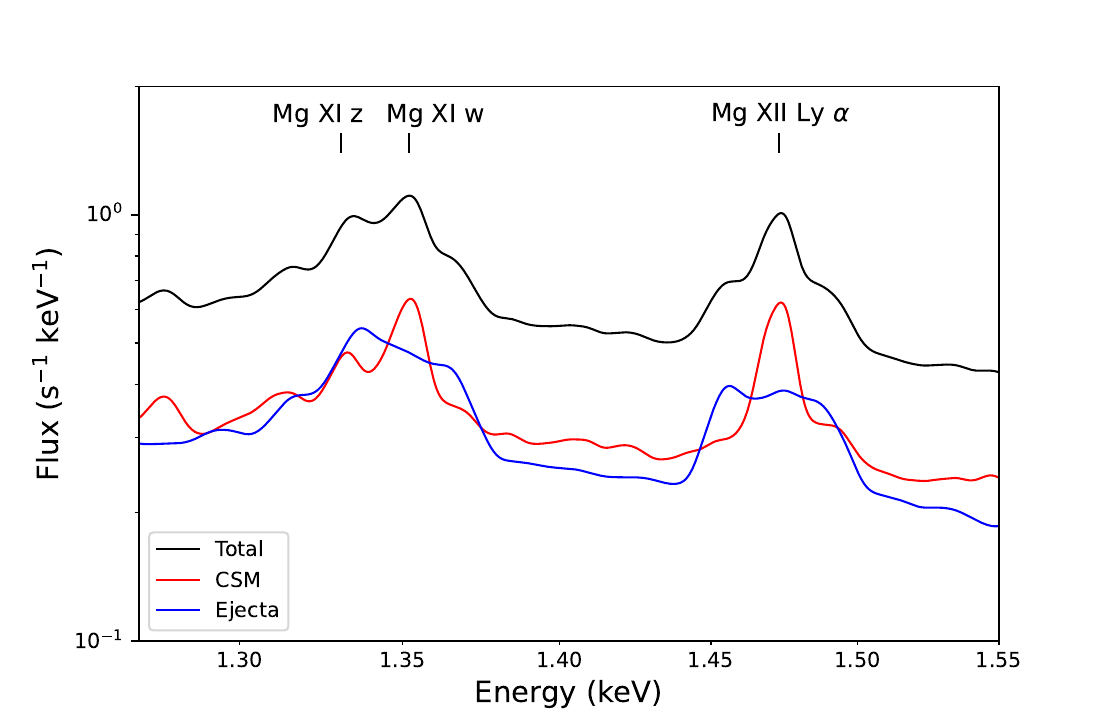}
    \includegraphics[width=\columnwidth]{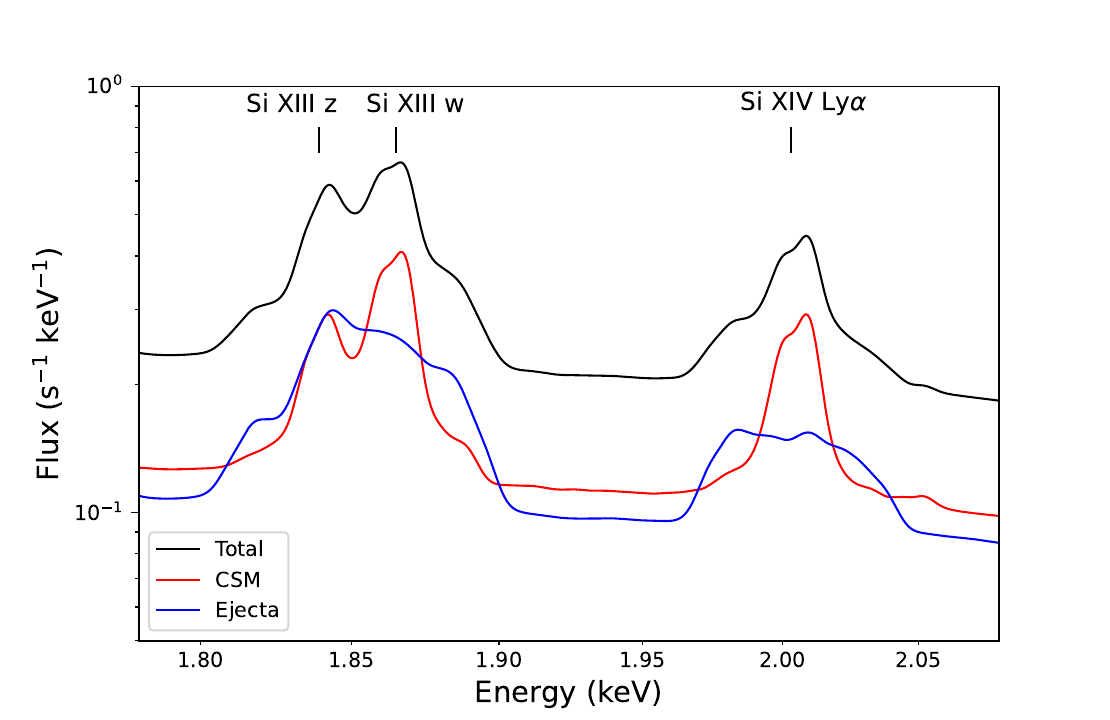}
    \caption{\textit{Upper panel:}Synthetic XRISM-Resolve spectra of SN 1987A for the year 2024 with an exposure time of 100 ks. 
    The orange spectrum was synthesized without any broadening effects except for the instrumental broadening, and it was multiplied by a factor 10 for visualization purposes. 
    The black spectrum was synthesized including bulk motion Doppler broadening. 
    All spectra are binned using the \cite{2016A&A...587A.151K} optimal binning algorithm. 
    The synthetic XRISM-Resolve spectral model is superimposed with a black dashed line. 
    The red curve shows the contribution of the CSM (ER and \hii) to the spectral model, while the blue curve shows the contribution of the ejecta.
    \textit{Lower left panel:} Close-up view of the spectral model shown in upper panel in the $1.27-1.55$ keV band.
    \textit{Lower right panel:} Same as left panel in the $1.78-2.08$ keV band.
    }
    \label{fig:spec}
\end{figure*}
As an initial step, we generate the XRISM-Resolve spectrum of SN 1987A, without applying any broadening effects, for the year 2024, to highlight the effects of broadening, shifts, and blending of emission lines by comparing this spectrum with the one including all broadening effects.
We assume an exposure time of 100 ks, mirroring the allocated time granted for the PV phase of the mission.
The resulting spectrum is presented in Fig. \ref{fig:spec} (upper panel, orange crosses).

Next, we introduce the broadening and shift effect caused by the bulk motion of the plasma, the black spectrum in Fig. \ref{fig:spec} (upper panel) illustrates the outcome of this operation.
The Doppler effect significantly broadens the emission lines, giving rise to complex profiles and line blending. 
These intricate and extensively broadened emission lines reflect the presence of different contributions from the different parts of the remnant in shaping the observed profiles.
Fig. \ref{fig:spec} also shows the total spectral model (black dashed curve) for the X-ray emission from SN 1987A, as seen by XRISM-Resolve, and its ejecta (blue curve) and CSM (red curve) component.
The contribution to the emission from the ejecta is comparable to the contribution from the CSM, as already noted in \cite{2020A&A...636A..22O}.

Fig. \ref{fig:spec} (lower panels) shows a close up view of the spectral model in the Mg (lower left panel) and Si (lower right panel) emission lines energy bands, respectively.
The Mg XII Ly$\alpha$ and the Si XIV Ly$\alpha$ lines, being quite isolated, are the perfect probe for examining the line profile.
The line profile, indeed, presents a composite structure with broad wings and a relatively narrow peak.
The broad line wings are clearly associated with the rapidly expanding ejecta, whose emission is marked by the blue curve in Fig. \ref{fig:spec}.
The narrow and slightly blue shifted central peak reflects the CSM contribution (see the red solid curve in Fig. \ref{fig:spec}) to the emission.
The Mg XI w, z and the Si XIII w, z emission lines, instead, exhibit a broader and blended profile, with a dominating influence of the ejecta in shaping the X-ray emission compared to the CSM.
We do not consider the Mg XI x, y and the Si XIII x, y lines because it can be seen from the spectrum without broadening (Fig. \ref{fig:spec}, upper panel, orange crosses) that they do not contribute significantly to the X-ray emission.
This pronounced effect can be primarily attributed to the ionization state of the ejecta, which are freshly shocked, and therefore heavily under-ionized (as shown in Fig. \ref{fig:ktvtau}). 
Consequently, their contribution to the emission of less ionized species, such as the Mg XI w, z and Si XIII w, z lines, is higher than in lines of H-like ions.

To point out the time evolution of the ejecta contribution to the X-ray emission predicted by our MHD model, we show in Fig. \ref{fig:ktvtau} (lower panels) the distribution of the EM of the X-ray emitting plasma as a function of the plasma velocity along the line of sight in 2011 and 2024.
The comparison between the 2011 and 2024 distributions underscores significant differences between these two epochs. 
The EM of the CSM has increased by a factor of $\sim2-3$, accompanied by a slight blue shift of its peak, due to the random distribution of dense clumps in the ring, which may have coincidentally placed one or two dense clumps in the blue-shifted portion of the ring.
In 2024, a large contribution to the line broadening arises from the freshly shocked and fast moving outer ejecta.
Indeed, the actual contribution to the X-ray emission from the ejecta in SN 1987A has been observed increasing year by year, as attested by \cite{2021ApJ...916...41S} and \cite{2022A&A...661A..30M}.
This is also clearly visible in Fig. \ref{fig:ktvtau} (lower panels), where the contribution of the ejecta has increased by about 2 orders of magnitude from the year 2011 to the year 2024.

\begin{figure*}[t]
    \centering
    \includegraphics[width=\columnwidth]{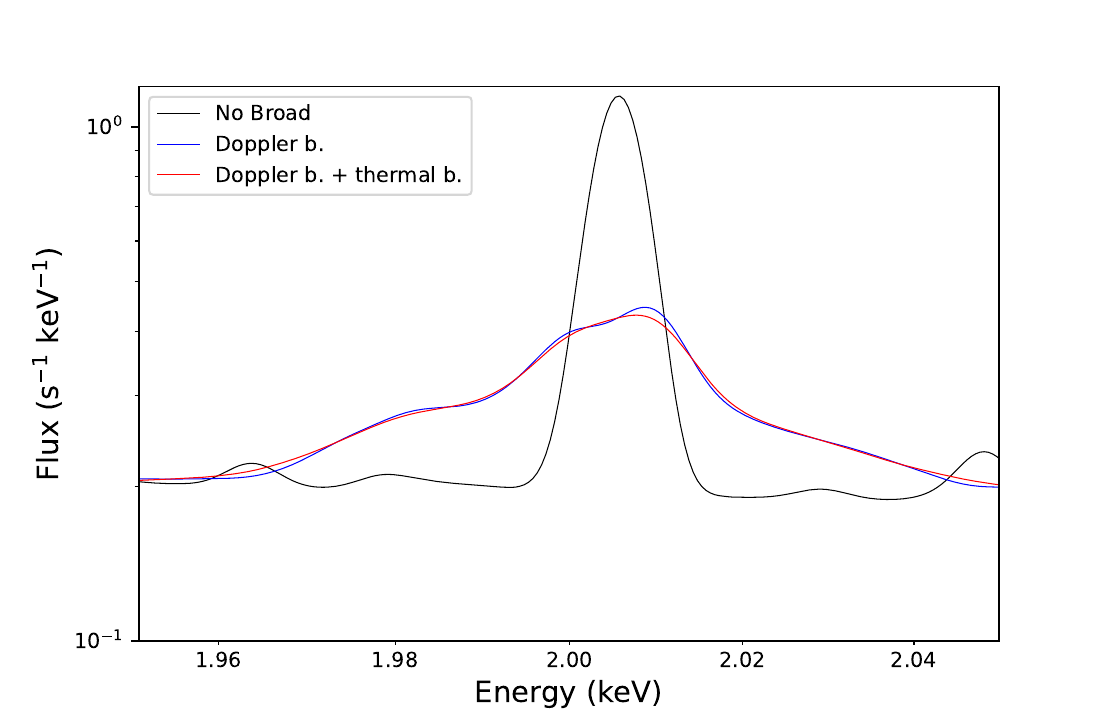}
    \includegraphics[width=\columnwidth]{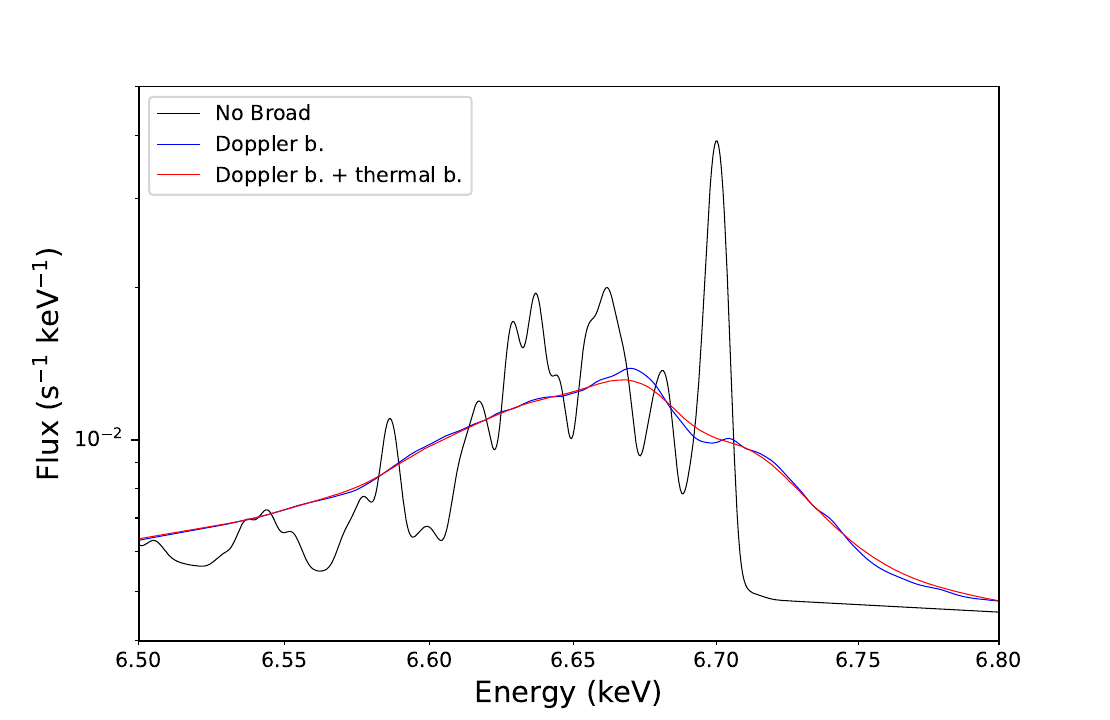}
    \caption{
    \textit{Left panel:} Close-up view of the synthetic X-ray spectral model of SN 1987A for the Si XIV Ly$\alpha$ emission line.
    The figure displays the spectrum synthesized in three ways: the black curve without including Doppler and thermal broadening, the blue curve synthesized by considering only Doppler broadening, and the red curve synthesized by accounting for both Doppler and thermal broadening.
    \textit{Right panel:} same as left panel for the Fe XXV emission lines complex.
    }
    \label{fig:Thermal}
\end{figure*}
We then synthesize the XRISM-Resolve spectrum of SN 1987A, adding the contribution of the thermal broadening to that of the Doppler broadening.
We here focus on two emission lines: the Si XIV  Ly$\alpha$ line to have a comparison with previous studies (\citealt{2019NatAs...3..236M}, \citealt{2023PPCF...65c4003M}) and the Fe XXV w, x, y, z lines (hereafter, Fe XXV lines complex), to explore the unprecedented capability of resolving the thermal broadening at high energy provided by this new instrument.
We do not discuss here the line emission of H-like iron (Fe XXVI  Ly$\alpha$), which could have been a much more simple probe for the thermal broadening, because it does not emerge from the continuum (see Fig. \ref{fig:spec}).
Fig. \ref{fig:Thermal} shows a comparison between the synthetic spectrum considering the Doppler broadening and the synthetic spectrum including also the thermal broadening, for the Si XIV Ly$\alpha$ line (left panel) and the Fe XXV lines complex (right panel).
From both the plots it is clear that, by adding the thermal broadening contribution to the spectrum, the shape of the emission line does not change significantly.
A direct comparison between the Si XIV Ly$\alpha$ emission lines in 2007, 2011 (\citealt{2019NatAs...3..236M}) and 2018 (\citealt{2021ApJ...922..140R,2023PPCF...65c4003M}) with those from this work, shows that in 2024 the Doppler broadening is expected to increase so much (because of the increased emission from the fast moving ejecta) that the relative contribution of thermal broadening is predicted to become negligible.

\begin{figure*}
    \centering
    \includegraphics[width=0.495\textwidth]{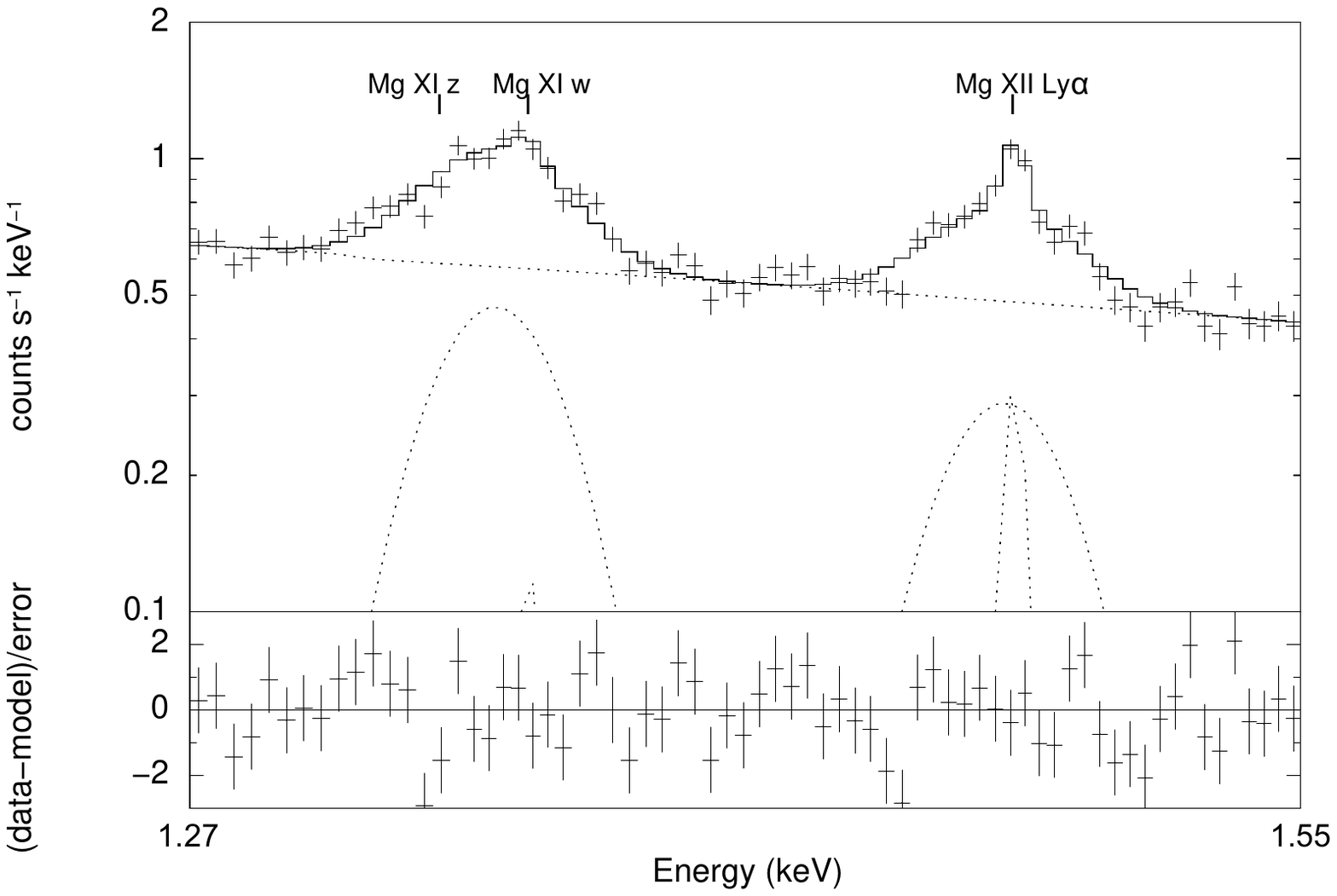} 
    \includegraphics[width=0.495\textwidth]{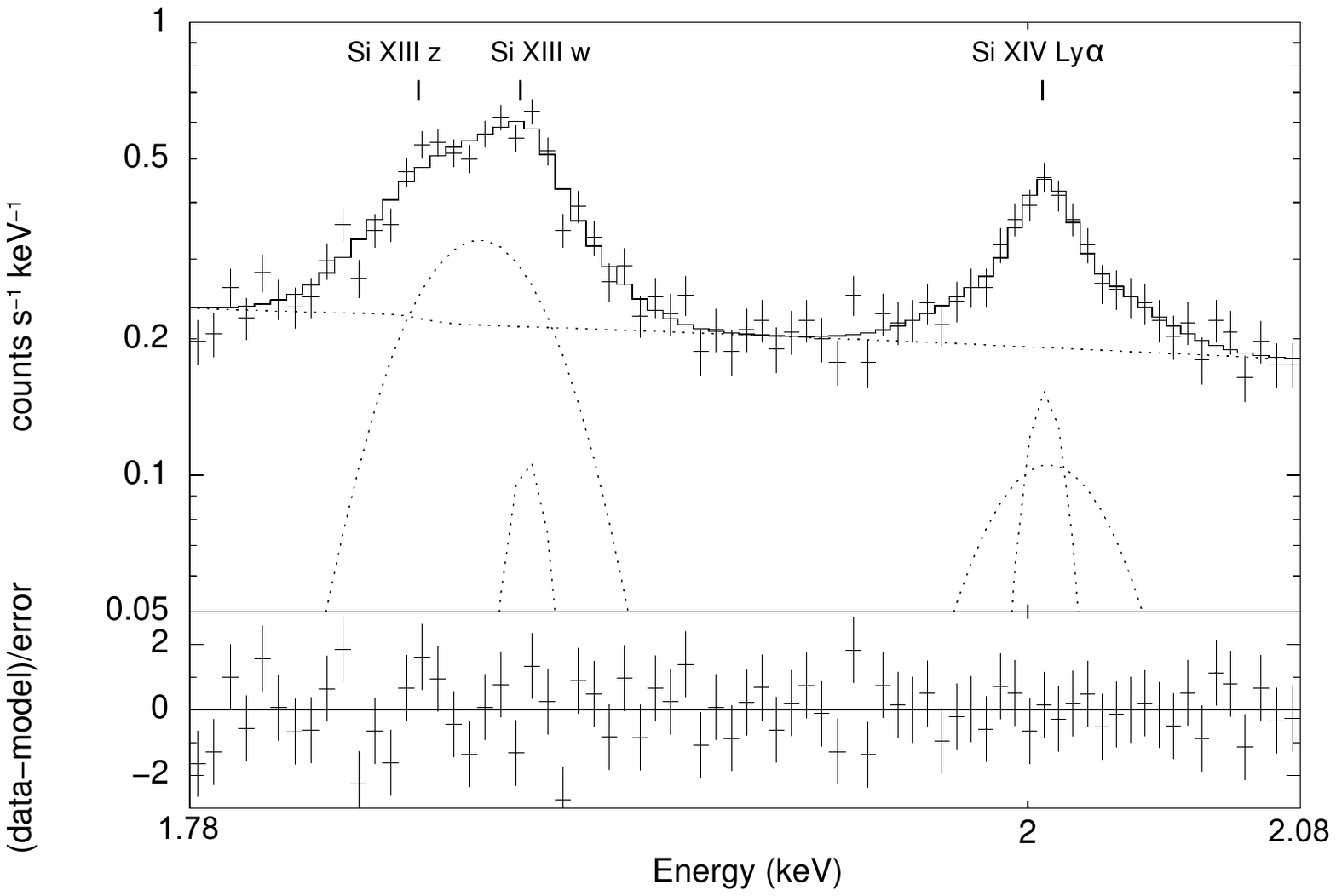}
    \caption{
    \textit{Left panel:} Close-up view of the synthetic  XRISM-Resolve spectrum (Fig. \ref{fig:spec}, black crosses)  highlighting the Mg XI w, z and Mg XII Ly$\alpha$ lines ($1.27-1.55$ keV band), with the corresponding best-fit model and residual. 
    \textit{Right panel:} Same as left panel for the Si XIII w, z and Si XIV Ly$\alpha$ lines ($1.78-2.08$ keV band).
    }
    \label{fig:speccomp}
\end{figure*}
Hereafter, we show how the effect of the ejecta dynamics can be revealed in the forthcoming 100 ks XRISM-Resolve observation of SN 1987A, which will unveil the role of the ejecta in shaping the line profiles shown in Fig. \ref{fig:spec}.
We adopt four Gaussian components, added to a thermal continuum (we also include the effect of interstellar and intergalactic absorption) to model the synthetic emission lines of Mg (Fig. \ref{fig:speccomp}, left panel) in the $1.27-1.55$ keV band. We follow the same approach to model the Si lines (Fig. \ref{fig:speccomp}, right panel) in the $1.78-2.08$ keV band.
We include these lines in our fitting procedure to accurately constrain the broadening of the lines stemming from the ejecta (which dominates the emission for He-like ions, see Fig. \ref{fig:spec}) and from the CSM (which are visible in the emission of H-like ions). 
Specifically, two Gaussian lines model the narrow CSM components respectively, in the lines of H-like ions (i.e. Mg XII Ly$\alpha$ and Si XIV Ly$\alpha$), and the He-like ions (i.e. Mg XI w and Si XIII w).
The other two Gaussian lines are adopted to model the broad ejecta components in the lines of He-like and H-like ions.
Given $\sigma_{3,1}$ and $E_{3,1}$ the width and the centroid of the ejecta components in the lines of He-like and H-like species, respectively, we impose $\sigma_1= \sigma_3\cdot E_1/E_3$ in the fitting procedure.
Similarly, we impose $\sigma_2= \sigma_4\cdot E_2/E_4$, being $\sigma_{4,2}$ and $E_{4,2}$ the width and the centroid of the CSM components in the lines of He-like and H-like species.

%The fitting procedure has been carried out making use of C-statistics.
Fig. \ref{fig:speccomp} shows the spectra of the Mg (left) and Si (right) lines with the corresponding best fit models and residuals.
The best fit values with error bars at the 90\% confidence level, are reported in Table \ref{tab:spec}.
The different components (\#1, \#3, \#4 in Table \ref{tab:spec}) are detected with high significance (their normalization being larger than zero at more than the 4$\sigma$ confidence level) and clearly show different expansion velocities for CSM and ejecta.
The \#2 lines in Table \ref{tab:spec} are detected with 2$\sigma$ confidence level for the Mg XI w and 3$\sigma$ confidence level for the Si XIII w confidence level.

We caution the reader that in the procedure described above we opt for a single Gaussian line to characterize the He-like ejecta triplets. 
The introduction of additional Gaussian lines for the He-like triplets is not statistically necessary and does not enhance the quality of the fit significantly. 
Nevertheless, we verify that the line broadenings reported in Table \ref{tab:spec} are not affected by this issue and are robust. To this end, we consider a spectral model including all the lines (i.e. three lines for the He-like triplets and one line for the H-like emission, for both the ejecta and CSM components). Though with the statistics provided by a 100 ks observation it is not possible to constrain such a large number of free parameters, we can determine the line centroids and normalizations from our simulation by considering a synthetic spectrum with an unrealistically high exposure time ($t_{exp}=10^8$ s). We then find that the line widths for the ejecta remain consistent within the error bars with those shown in Table \ref{tab:spec}, which then provide reliable values for the velocity of ejecta and  CSM.

\begin{table}[t!]
    \caption{Best-ft values for the synthetic spectrum in the Mg and Si energy bands and the velocities retrieved from the Doppler broadening. Error bars are at 90\% confidence level.}
    \resizebox{\columnwidth}{!}{
    \begin{tabular}{cccccc}
    \hline\hline

        \#&Emission line&Energy (keV)&$\sigma$ (eV)&Velocity (km s$^{-1}$)&Norm (cm$^{-2}$ s$^{-1}$)\\
        \hline
         \multicolumn{6}{c}{Mg $C-stat/d.o.f.=85.09/61$}\\
         1&Mg XI w, z& 1.3440$_{-0.0014}^{+0.0018}$& & & 8.0$_{-0.7}^{+0.6}\times10^{-5}$ \\
         2&Mg XI w& 1.352$_{-0.007}^{+0.007}$& & & 0.4$_{-0.3}^{+0.3}\times10^{-5}$ \\
         3&Mg XII Ly$\alpha$& 1.4709$_{-0.0022}^{+0.0017}$& 17.7$_{-0.7}^{+1.2}$& $3610_{-140}^{+240}$& 4.5$_{-0.5}^{+0.5}\times10^{-5}$ \\
         4&Mg XII Ly$\alpha$ & 1.4734$_{-0.0010}^{+0.0010}$& $<3$ & $<600$ & 0.9$_{-0.2}^{+0.2}\times10^{-5}$ \\
         \multicolumn{6}{c}{Si $C-stat/d.o.f.=62.79/63$}\\
         1&Si XIII w, z& 1.8543$_{-0.0015}^{+0.0014}$& & & 6.9$_{-0.5}^{+0.5}\times10^{-5}$ \\
         2&Si XIII w& 1.867$_{-0.004}^{+0.002}$& & &0.7$_{-0.3}^{+0.4}\times10^{-5}$ \\
         3&Si XIV Ly$\alpha$& 2.005$_{-0.004}^{+0.005}$   & 21.1$_{-1.3}^{+1.4}$& $3160_{-190}^{+210}$& 2.2$_{-0.6}^{+0.4}\times10^{-5}$ \\
         4&Si XIV Ly$\alpha$& 2.0047$_{-0.0020}^{+0.0019}$   & 5.3$_{-1.5}^{+1.9}$       & $800_{-200}^{+300}$ & 1.1$_{-0.3}^{+0.5}\times10^{-5}$ \\
         \hline
    \end{tabular}
    }
    \label{tab:spec}
\end{table}

\section{Discussion and Conclusions}\label{sect:con}
In this letter, we present a spectral synthesis of the X-ray emission of SN 1987A, mirroring the new XRISM-Resolve micro-calorimeter observation, which will take place during the PV phase of the mission, shedding new light on this intriguing celestial object.
We used a recently proposed approach, which leverages 3D MHD modelling to derive  observables for different epochs, taking into account self-consistently the evolution of the remnant since the core-collapse.
The resulting synthetic spectrum, which takes into account the effect of the Doppler shifts and broadening due to the bulk motion of the X-ray emitting plasma along the line of sight, shows a complex and largely broadened profile for the emission lines.

The model used in this study indicates that X-ray emission was primarily dominated by the shocked CSM, with a smaller contribution from shocked ejecta in epochs prior to 2020.
This resulted in relatively narrow emission lines, reflecting the bulk motion of shocked CSM material (mainly from the ER) along the line of sight (\citealt{2019NatAs...3..236M}, \citealt{2021ApJ...922..140R}, \citealt{2023PPCF...65c4003M}). 
We note that the emission lines predicted here for 2024 exhibit a significantly larger broadening than before.
This is due to the increased contribution from shocked ejecta, which are characterized by higher velocities along the line of sight.
Indeed, in 2024 the contribution to the X-ray emission stemming from freshly-shocked and fast moving ejecta will be two orders of magnitude higher than in 2011 (see Fig. \ref{fig:ktvtau}, lower panels).
The increased Doppler broadening of the lines will provide information on the ejecta velocity (though hampering the measurement of ion temperatures).

In our model, the X-ray emission from the ejecta originates from the metal-poor outer envelope. We note that \cite{2023ApJ...949L..27L} recently analyzed JWST NIRSpec observations of SN 1987A, finding evidence for two Fe-rich inner ejecta plumes interacting with the reverse shock, not predicted by the model we adopted.
This fast moving material could increase the emissivity of the Fe lines of the actual spectrum with respect to our predictions, thus further enhancing the broadening effect in these lines.

In any case, the measurement of the Doppler broadening will provide direct evidence for the shocked ejecta and their expansion.  
We have shown in Sect. \ref{sect:res} that the forthcoming 100 ks XRISM-Resolve observation of SN 1987A will allow us to derive the ejecta dynamics by disentangling the peaked emission of the CSM from the broad wings stemming from the ejecta in the emission lines of H-like ions, and simultaneously measuring the Doppler broadening from the emission lines of He-like ions, dominated by the ejecta emission. In particular, from the fitting of the synthetic Mg and Si line profiles (obtained with an exposure time of 100 ks) we retrieved an ejecta velocity of $3610_{-140}^{+240}$ km s$^{-1}$ and $3160_{-190}^{+210}$ km s$^{-1}$ (see Table \ref{tab:spec}). We notice that the ``true" values (directly desumed from the MHD simulation) are $3480$ km s$^{-1}$ and $3280$ km s$^{-1}$, respectively, thus confirming that synthetic spectra provide a very robust diagnostic tool. 
We point out that, since emission depends on temperature, ionization parameter and particle density, we do not expect the broadening of Mg and Si lines to be exactly the same.
Also for the same reason, we do not expect the ratio between ejecta and CSM for the integrated flux of the Ly$\alpha$ lines to be the same in Mg and Si lines.

In summary, our findings represent an exploration of the spectral characteristics of SN 1987A, leveraging the capabilities of the XRISM-Resolve micro-calorimeter spectrometer. 
Our analysis sheds light on the diverse contributions from the CSM and ejecta in the next observations from the new satellite.
We also emphasize the relevance of Doppler broadening in shaping the emission lines for future observations. 
These insights lay the groundwork for future observations and provide a deeper understanding of the evolving dynamics within this iconic supernova remnant.
\vspace{8pt}

{\noindent \small \textit{Acknowledgments} - This work was supported by JSPS Core-to-Core Program, (grant number:JPJSCCA20220002).
This work was financially supported by Japan Society for the Promotion of Science Grants-in-Aid for Scientific Research (KAKENHI) Grant Number,  JP23H01211 (AB), JP20H00174 (S.K.) and JP21H01121 (S.K. and Y.T.).
V.S. acknowledges the financial support received from the University of Palermo. 
MM, SO, and FB acknowledge financial contribution from the PRIN MUR "Life, death and after-death of massive stars: reconstructing the path from the pre-supernova evolution to the supernova remnant" and the Astrofund Theory Grant of INAF.
S.N. was supported by JSPS KAKENHI (A) Grant Number JP19H00693 and RIKEN Pioneering Project for Evolution of Matter in the Universe (r-EMU).}

\bibliography{sample631}{}

\begin{thebibliography}{}
\expandafter\ifx\csname natexlab\endcsname\relax\def\natexlab#1{#1}\fi
\providecommand{\url}[1]{\href{#1}{#1}}
\providecommand{\dodoi}[1]{doi:~\href{http://doi.org/#1}{\nolinkurl{#1}}}
\providecommand{\doeprint}[1]{\href{http://ascl.net/#1}{\nolinkurl{http://ascl.net/#1}}}
\providecommand{\doarXiv}[1]{\href{https://arxiv.org/abs/#1}{\nolinkurl{https://arxiv.org/abs/#1}}}

\bibitem[{{Abell{\'a}n} {et~al.}(2017){Abell{\'a}n}, {Indebetouw}, {Marcaide},
  {Gabler}, {Fransson}, {Spyromilio}, {Burrows}, {Chevalier}, {Cigan},
  {Gaensler}, {Gomez}, {Janka}, {Kirshner}, {Larsson}, {Lundqvist}, {Matsuura},
  {McCray}, {Ng}, {Park}, {Roche}, {Staveley-Smith}, {van Loon}, {Wheeler}, \&
  {Woosley}}]{2017ApJ...842L..24A}
{Abell{\'a}n}, F.~J., {Indebetouw}, R., {Marcaide}, J.~M., {et~al.} 2017,
  \apjl, 842, L24, \dodoi{10.3847/2041-8213/aa784c}

\bibitem[{{Arnaud}(1996)}]{1996ASPC..101...17A}
{Arnaud}, K.~A. 1996, in Astronomical Society of the Pacific Conference Series,
  Vol. 101, Astronomical Data Analysis Software and Systems V, ed. G.~H.
  {Jacoby} \& J.~{Barnes}, 17

\bibitem[{{Beuermann} {et~al.}(1994){Beuermann}, {Brandt}, \&
  {Pietsch}}]{1994A&A...281L..45B}
{Beuermann}, K., {Brandt}, S., \& {Pietsch}, W. 1994, \aap, 281, L45

\bibitem[{{Boggs} {et~al.}(2015){Boggs}, {Harrison}, {Miyasaka},
  {Grefenstette}, {Zoglauer}, {Fryer}, {Reynolds}, {Alexander}, {An}, {Barret},
  {Christensen}, {Craig}, {Forster}, {Giommi}, {Hailey}, {Hornstrup},
  {Kitaguchi}, {Koglin}, {Madsen}, {Mao}, {Mori}, {Perri}, {Pivovaroff},
  {Puccetti}, {Rana}, {Stern}, {Westergaard}, \& {Zhang}}]{2015Sci...348..670B}
{Boggs}, S.~E., {Harrison}, F.~A., {Miyasaka}, H., {et~al.} 2015, Science, 348,
  670, \dodoi{10.1126/science.aaa2259}

\bibitem[{{Borkowski} {et~al.}(1997){Borkowski}, {Blondin}, \&
  {McCray}}]{1997ApJ...477..281B}
{Borkowski}, K.~J., {Blondin}, J.~M., \& {McCray}, R. 1997, \apj, 477, 281,
  \dodoi{10.1086/303691}

\bibitem[{{Dotani} {et~al.}(1987){Dotani}, {Hayashida}, {Inoue}, {Itoh},
  {Koyama}, {Makino}, {Mitsuda}, {Murakami}, {Oda}, {Ogawara}, {Takano},
  {Tanaka}, {Yoshida}, {Makishima}, {Ohashi}, {Kawai}, {Matsuoka}, {Hoshi},
  {Hayakawa}, {Kii}, {Kunieda}, {Nagase}, {Tawara}, {Hatsukade}, {Kitamoto},
  {Miyamoto}, {Tsunemi}, {Yamashita}, {Nakagawa}, {Yamauchi}, {Turner},
  {Pounds}, {Thomas}, {Stewart}, {Cruise}, {Patchett}, \&
  {Reading}}]{1987Natur.330..230D}
{Dotani}, T., {Hayashida}, K., {Inoue}, H., {et~al.} 1987, \nat, 330, 230,
  \dodoi{10.1038/330230a0}

\bibitem[{{Frank} {et~al.}(2016){Frank}, {Zhekov}, {Park}, {McCray}, {Dwek}, \&
  {Burrows}}]{2016ApJ...829...40F}
{Frank}, K.~A., {Zhekov}, S.~A., {Park}, S., {et~al.} 2016, \apj, 829, 40,
  \dodoi{10.3847/0004-637X/829/1/40}

\bibitem[{{Greco} {et~al.}(2020){Greco}, {Vink}, {Miceli}, {Orlando},
  {Dom{\v{c}}ek}, {Zhou}, {Bocchino}, \& {Peres}}]{2020A&A...638A.101G}
{Greco}, E., {Vink}, J., {Miceli}, M., {et~al.} 2020, \aap, 638, A101,
  \dodoi{10.1051/0004-6361/202038092}

\bibitem[{{Greco} {et~al.}(2021){Greco}, {Miceli}, {Orlando}, {Olmi},
  {Bocchino}, {Nagataki}, {Ono}, {Dohi}, \& {Peres}}]{2021ApJ...908L..45G}
{Greco}, E., {Miceli}, M., {Orlando}, S., {et~al.} 2021, \apjl, 908, L45,
  \dodoi{10.3847/2041-8213/abdf5a}

\bibitem[{{Greco} {et~al.}(2022){Greco}, {Miceli}, {Orlando}, {Olmi},
  {Bocchino}, {Nagataki}, {Sun}, {Vink}, {Sapienza}, {Ono}, {Dohi}, \&
  {Peres}}]{2022ApJ...931..132G}
---. 2022, \apj, 931, 132, \dodoi{10.3847/1538-4357/ac679d}

\bibitem[{{Haas} {et~al.}(1990){Haas}, {Colgan}, {Erickson}, {Lord}, {Burton},
  \& {Hollenbach}}]{1990ApJ...360..257H}
{Haas}, M.~R., {Colgan}, S. W.~J., {Erickson}, E.~F., {et~al.} 1990, \apj, 360,
  257, \dodoi{10.1086/169115}

\bibitem[{{Haberl} {et~al.}(2006){Haberl}, {Geppert}, {Aschenbach}, \&
  {Hasinger}}]{2006A&A...460..811H}
{Haberl}, F., {Geppert}, U., {Aschenbach}, B., \& {Hasinger}, G. 2006, \aap,
  460, 811, \dodoi{10.1051/0004-6361:20066198}

\bibitem[{{Ishisaki} {et~al.}(2022){Ishisaki}, {Kelley}, {Awaki}, {Balleza},
  {Barnstable}, {Bialas}, {Boissay-Malaquin}, {Brown}, {Canavan}, {Cumbee},
  {Carnahan}, {Chiao}, {Comber}, {Costantini}, {den Herder}, {Dercksen}, {de
  Vries}, {DiPirro}, {Eckart}, {Ezoe}, {Ferrigno}, {Fujimoto}, {Gorter},
  {Graham}, {Grim}, {Hartz}, {Hayakawa}, {Hayashi}, {Hell}, {Hoshino},
  {Ichinohe}, {Ishida}, {Ishikawa}, {James}, {Kenyon}, {Kilbourne}, {Kimball},
  {Kitamoto}, {Leutenegger}, {Maeda}, {McCammon}, {Miko}, {Mizumoto},
  {Okajima}, {Okamoto}, {Paltani}, {Porter}, {Sato}, {Sato}, {Sawada},
  {Shinozaki}, {Shipman}, {Shirron}, {Sneiderman}, {Soong}, {Szymkiewicz},
  {Szymkowiak}, {Takei}, {Tamura}, {Tsujimoto}, {Uchida}, {Wasserzug},
  {Witthoeft}, {Wolfs}, {Yamada}, \& {Yasuda}}]{2022SPIE12181E..1SI}
{Ishisaki}, Y., {Kelley}, R.~L., {Awaki}, H., {et~al.} 2022, in Society of
  Photo-Optical Instrumentation Engineers (SPIE) Conference Series, Vol. 12181,
  Space Telescopes and Instrumentation 2022: Ultraviolet to Gamma Ray, ed.
  J.-W.~A. {den Herder}, S.~{Nikzad}, \& K.~{Nakazawa}, 121811S,
  \dodoi{10.1117/12.2630654}

\bibitem[{{Ivanova} {et~al.}(2002){Ivanova}, {Podsiadlowski}, \&
  {Spruit}}]{2002MNRAS.334..819I}
{Ivanova}, N., {Podsiadlowski}, P., \& {Spruit}, H. 2002, \mnras, 334, 819,
  \dodoi{10.1046/j.1365-8711.2002.05543.x}

\bibitem[{{Kaastra} \& {Bleeker}(2016)}]{2016A&A...587A.151K}
{Kaastra}, J.~S., \& {Bleeker}, J.~A.~M. 2016, \aap, 587, A151,
  \dodoi{10.1051/0004-6361/201527395}

\bibitem[{{Larsson} {et~al.}(2023){Larsson}, {Fransson}, {Sargent}, {Jones},
  {Barlow}, {Bouchet}, {Meixner}, {Blommaert}, {Coulais}, {Fox}, {Gastaud},
  {Glasse}, {Habel}, {Hirschauer}, {Hjorth}, {Jaspers}, {Kavanagh}, {Krause},
  {Lau}, {Lenki{\'c}}, {Nayak}, {Rest}, {Temim}, {Tikkanen}, {Wesson}, \&
  {Wright}}]{2023ApJ...949L..27L}
{Larsson}, J., {Fransson}, C., {Sargent}, B., {et~al.} 2023, \apjl, 949, L27,
  \dodoi{10.3847/2041-8213/acd555}

\bibitem[{{Maggi} {et~al.}(2012){Maggi}, {Haberl}, {Sturm}, \&
  {Dewey}}]{2012A&A...548L...3M}
{Maggi}, P., {Haberl}, F., {Sturm}, R., \& {Dewey}, D. 2012, \aap, 548, L3,
  \dodoi{10.1051/0004-6361/201220595}

\bibitem[{{Maitra} {et~al.}(2022){Maitra}, {Haberl}, {Sasaki}, {Maggi},
  {Dennerl}, \& {Freyberg}}]{2022A&A...661A..30M}
{Maitra}, C., {Haberl}, F., {Sasaki}, M., {et~al.} 2022, \aap, 661, A30,
  \dodoi{10.1051/0004-6361/202141104}

\bibitem[{{McCray} \& {Fransson}(2016)}]{2016ARA&A..54...19M}
{McCray}, R., \& {Fransson}, C. 2016, \araa, 54, 19,
  \dodoi{10.1146/annurev-astro-082615-105405}

\bibitem[{{Miceli}(2023)}]{2023PPCF...65c4003M}
{Miceli}, M. 2023, Plasma Physics and Controlled Fusion, 65, 034003,
  \dodoi{10.1088/1361-6587/acb082}

\bibitem[{{Miceli} {et~al.}(2019){Miceli}, {Orlando}, {Burrows}, {Frank},
  {Argiroffi}, {Reale}, {Peres}, {Petruk}, \& {Bocchino}}]{2019NatAs...3..236M}
{Miceli}, M., {Orlando}, S., {Burrows}, D.~N., {et~al.} 2019, Nature Astronomy,
  3, 236, \dodoi{10.1038/s41550-018-0677-8}

\bibitem[{{Ono} {et~al.}(2020){Ono}, {Nagataki}, {Ferrand}, {Takahashi},
  {Umeda}, {Yoshida}, {Orlando}, \& {Miceli}}]{2020ApJ...888..111O}
{Ono}, M., {Nagataki}, S., {Ferrand}, G., {et~al.} 2020, \apj, 888, 111,
  \dodoi{10.3847/1538-4357/ab5dba}

\bibitem[{{Orlando} {et~al.}(2015){Orlando}, {Miceli}, {Pumo}, \&
  {Bocchino}}]{2015ApJ...810..168O}
{Orlando}, S., {Miceli}, M., {Pumo}, M.~L., \& {Bocchino}, F. 2015, \apj, 810,
  168, \dodoi{10.1088/0004-637X/810/2/168}

\bibitem[{{Orlando} {et~al.}(2019){Orlando}, {Miceli}, {Petruk}, {Ono},
  {Nagataki}, {Aloy}, {Mimica}, {Lee}, {Bocchino}, {Peres}, \&
  {Guarrasi}}]{2019A&A...622A..73O}
{Orlando}, S., {Miceli}, M., {Petruk}, O., {et~al.} 2019, \aap, 622, A73,
  \dodoi{10.1051/0004-6361/201834487}

\bibitem[{{Orlando} {et~al.}(2020){Orlando}, {Ono}, {Nagataki}, {Miceli},
  {Umeda}, {Ferrand}, {Bocchino}, {Petruk}, {Peres}, {Takahashi}, \&
  {Yoshida}}]{2020A&A...636A..22O}
{Orlando}, S., {Ono}, M., {Nagataki}, S., {et~al.} 2020, \aap, 636, A22,
  \dodoi{10.1051/0004-6361/201936718}

\bibitem[{{Panagia}(1999)}]{1999IAUS..190..549P}
{Panagia}, N. 1999, in New Views of the Magellanic Clouds, ed. Y.~H. {Chu},
  N.~{Suntzeff}, J.~{Hesser}, \& D.~{Bohlender}, Vol. 190, 549

\bibitem[{{Park} {et~al.}(2006){Park}, {Zhekov}, {Burrows}, {Garmire},
  {Racusin}, \& {McCray}}]{2006ApJ...646.1001P}
{Park}, S., {Zhekov}, S.~A., {Burrows}, D.~N., {et~al.} 2006, \apj, 646, 1001,
  \dodoi{10.1086/505023}

\bibitem[{{Park} {et~al.}(2005){Park}, {Zhekov}, {Burrows}, \&
  {McCray}}]{2005ApJ...634L..73P}
{Park}, S., {Zhekov}, S.~A., {Burrows}, D.~N., \& {McCray}, R. 2005, \apjl,
  634, L73, \dodoi{10.1086/498848}

\bibitem[{{Ravi} {et~al.}(2021){Ravi}, {Park}, {Zhekov}, {Miceli}, {Orlando},
  {Frank}, \& {Burrows}}]{2021ApJ...922..140R}
{Ravi}, A.~P., {Park}, S., {Zhekov}, S.~A., {et~al.} 2021, \apj, 922, 140,
  \dodoi{10.3847/1538-4357/ac249a}

\bibitem[{{Sugerman} {et~al.}(2005){Sugerman}, {Crotts}, {Kunkel}, {Heathcote},
  \& {Lawrence}}]{2005ApJS..159...60S}
{Sugerman}, B. E.~K., {Crotts}, A. P.~S., {Kunkel}, W.~E., {Heathcote}, S.~R.,
  \& {Lawrence}, S.~S. 2005, \apjs, 159, 60, \dodoi{10.1086/430408}

\bibitem[{{Sun} {et~al.}(2021){Sun}, {Vink}, {Chen}, {Zhou}, {Prokhorov},
  {P{\"u}hlhofer}, \& {Malyshev}}]{2021ApJ...916...41S}
{Sun}, L., {Vink}, J., {Chen}, Y., {et~al.} 2021, \apj, 916, 41,
  \dodoi{10.3847/1538-4357/ac033d}

\bibitem[{{Tashiro} {et~al.}(2020){Tashiro}, {Maejima}, {Toda}, {Kelley},
  {Reichenthal}, {Hartz}, {Petre}, {Williams}, {Guainazzi}, {Costantini},
  {Fujimoto}, {Hayashida}, {Henegar-Leon}, {Holland}, {Ishisaki}, {Kilbourne},
  {Loewenstein}, {Matsushita}, {Mori}, {Okajima}, {Porter}, {Sneiderman},
  {Takei}, {Terada}, {Tomida}, {Yamaguchi}, {Watanabe}, {Akamatsu}, {Arai},
  {Audard}, {Awaki}, {Babyk}, {Bamba}, {Bando}, {Behar}, {Bialas},
  {Boissay-Malaquin}, {Brenneman}, {Brown}, {Canavan}, {Chiao}, {Comber},
  {Corrales}, {Cumbee}, {de Vries}, {den Herder}, {Dercksen}, {Diaz-Trigo},
  {DiPirro}, {Done}, {Dotani}, {Ebisawa}, {Eckart}, {Eckert}, {Eguchi},
  {Enoto}, {Ezoe}, {Ferrigno}, {Fujita}, {Fukazawa}, {Furuzawa}, {Gallo},
  {Gorter}, {Grim}, {Gu}, {Hagino}, {Hamaguchi}, {Hatsukade}, {Hawthorn},
  {Hayashi}, {Hell}, {Hiraga}, {Hodges-Kluck}, {Horiuchi}, {Hornschemeier},
  {Hoshino}, {Ichinohe}, {Iga}, {Iizuka}, {Ishida}, {Ishihama}, {Ishikawa},
  {Ishimura}, {Jaffe}, {Kaastra}, {Kallman}, {Kara}, {Katsuda}, {Kenyon},
  {Kimball}, {Kitaguchi}, {Kitamoto}, {Kobayashi}, {Kobayashi}, {Kohmura},
  {Kubota}, {Leutenegger}, {Li}, {Lockard}, {Maeda}, {Markevitch}, {Martz},
  {Matsumoto}, {Matsuzaki}, {McCammon}, {McLaughlin}, {McNamara}, {Miko},
  {Miller}, {Miller}, {Minesugi}, {Mitani}, {Mitsuishi}, {Mizumoto}, {Mizuno},
  {Mukai}, {Murakami}, {Mushotzky}, {Nakajima}, {Nakamura}, {Nakazawa},
  {Natsukari}, {Nigo}, {Nishioka}, {Nobukawa}, {Nobukawa}, {Noda}, {Odaka},
  {Ogawa}, {Ohashi}, {Ohno}, {Ohta}, {Okamoto}, {Ota}, {Ozaki}, {Paltani},
  {Plucinsky}, {Pottschmidt}, {Sampson}, {Sasaki}, {Sato}, {Sato}, {Sato},
  {Sawada}, {Seta}, {Shibano}, {Shida}, {Shidatsu}, {Shigeto}, {Shinozaki},
  {Shirron}, {Simionescu}, {Smith}, {Someya}, {Soong}, {Sugawara}, {Sugawara},
  {Szymkowiak}, {Takahashi}, {Takeshima}, {Tamagawa}, {Tamura}, {Tanaka},
  {Tanimoto}, {Terashima}, {Tsuboi}, {Tsujimoto}, {Tsunemi}, {Tsuru}, {Uchida},
  {Uchida}, {Uchiyama}, {Ueda}, {Uno}, {Vink}, {Watanabe}, {Witthoeft},
  {Wolfs}, {Yamada}, {Yamaoka}, {Yamasaki}, {Yamauchi}, {Yamauchi}, {Yanagase},
  {Yaqoob}, {Yasuda}, {Yoshida}, {Yoshioka}, \&
  {Zhuravleva}}]{2020SPIE11444E..22T}
{Tashiro}, M., {Maejima}, H., {Toda}, K., {et~al.} 2020, in Society of
  Photo-Optical Instrumentation Engineers (SPIE) Conference Series, Vol. 11444,
  Space Telescopes and Instrumentation 2020: Ultraviolet to Gamma Ray, ed.
  J.-W.~A. {den Herder}, S.~{Nikzad}, \& K.~{Nakazawa}, 1144422,
  \dodoi{10.1117/12.2565812}

\bibitem[{{Urushibata} {et~al.}(2018){Urushibata}, {Takahashi}, {Umeda}, \&
  {Yoshida}}]{2018MNRAS.473L.101U}
{Urushibata}, T., {Takahashi}, K., {Umeda}, H., \& {Yoshida}, T. 2018, \mnras,
  473, L101, \dodoi{10.1093/mnrasl/slx166}

\bibitem[{{West} {et~al.}(1987){West}, {Lauberts}, {Jorgensen}, \&
  {Schuster}}]{1987A&A...177L...1W}
{West}, R.~M., {Lauberts}, A., {Jorgensen}, H.~E., \& {Schuster}, H.~E. 1987,
  \aap, 177, L1

\bibitem[{{Zhekov} {et~al.}(2009){Zhekov}, {McCray}, {Dewey}, {Canizares},
  {Borkowski}, {Burrows}, \& {Park}}]{2009ApJ...692.1190Z}
{Zhekov}, S.~A., {McCray}, R., {Dewey}, D., {et~al.} 2009, \apj, 692, 1190,
  \dodoi{10.1088/0004-637X/692/2/1190}

\end{thebibliography}
\bibliographystyle{aasjournal}
\end{document}